\documentclass[preprint,prd,nofootinbib,tightenlines,amsmath]{revtex4}
\usepackage{bm}
\usepackage{epsf}
\usepackage{color}
\usepackage{graphicx}

\begin{document}
\baselineskip=17pt
\parskip=5pt

\hfill hep-ph/0506067\\
\hspace*{\fill} WSU-HEP-0504

\bigskip\bigskip\bigskip

\title{The Decay $\bm{\Sigma^{+}\to p \ell^{+}\ell^{-}}$ within the Standard Model}

\author{Xiao-Gang He}
\email{hexg@phys.ntu.edu.tw}
\affiliation{Department of Physics, National Taiwan University, Taipei}

\author{Jusak Tandean}
\email{jtandean@physics.wayne.edu}
\affiliation{Department of Physics and Astronomy, Wayne State University, Detroit, MI 48201}

\author{G. Valencia}
\email{valencia@iastate.edu} \affiliation{Department of Physics and
Astronomy, Iowa State University, Ames, IA 50011 \vspace{3ex} \\}

\date{\today}

\begin{abstract}

The HyperCP collaboration has recently reported the observation of three
events for the decay  $\Sigma^{+}\to p \mu^{+}\mu^{-}$. They have suggested
that new physics may be required to understand the implied decay rate and
the observed $M_{\mu\mu}^{}$ distribution.
Motivated by this result, we re-examine this mode within the standard model,
considering both the short-distance and long-distance contributions.
The long-distance part depends on four complex form-factors. We determine
their imaginary parts from unitarity, fix two of the real parts from
the $\Sigma^{+}\to p \gamma$ measurements, and estimate the other two with
vector-meson-dominance models.
Taking into account constraints from  $\Sigma^{+}\to p e^{+}e^{-}$,
we find that  $\Sigma^{+}\to p \mu^{+}\mu^{-}$  is long-distance dominated and
its rate falls within the range suggested by the HyperCP measurement.

\end{abstract}

\pacs{PACS numbers: }

\maketitle

\section{Introduction}

Three events for the decay mode \,$\Sigma^{+}\to p
\mu^{+}\mu^{-}$\, have been recently observed by the HyperCP (E871)
collaboration~\cite{Park:2005ek} with results that suggest new
physics may be needed to explain them. In this paper we re-examine
this mode~\cite{Bergstrom:1987wr} within the standard model.

There are short- and long-distance contributions to this decay. In
the standard model~(SM), the leading short-distance contribution
comes from the $Z$-penguin and box diagrams, as well as the electromagnetic
penguin with the photon connected to the dimuon pair~\cite{Buchalla:1995vs}.
We find that this contribution yields a branching ratio of order~$10^{-12}$,
which is much smaller than the central experimental value of
\,$8.6\times 10^{-8}$\,  reported by HyperCP~\cite{Park:2005ek}.
It is well known that the long-distance contribution to the weak radiative
mode \,$\Sigma^+\to p\gamma$\, is much larger than the short-distance contribution.
It is therefore also possible to have enhanced long-distance contributions to
\,$\Sigma^{+}\to p\mu^{+}\mu^{-}$\,  via an intermediate virtual photon from
\,$\Sigma^+\to p\gamma$.\,  We find that the resulting branching ratio is
in agreement with the measured value.  There is, of course, still
the possibility~\cite{He:1999ik} that new physics is responsible for
the observed branching ratio of  \,$\Sigma^+\to p\gamma$ and  hence
that of \,$\Sigma^{+}\to p\mu^{+}\mu^{-}$.\,
This implies that it is essential to have an up-to-date estimate of
the standard-model contributions, on which we concentrate in this work.

In Sec.~\ref{sd} we  update the estimate of the short-distance
amplitude. We use the standard effective Hamiltonian for the
\,$s\to d\ell^{+}\ell^{-}$\, transition~\cite{Buchalla:1995vs}
supplemented with hadronic matrix elements for the relevant
currents.
%We find that the short-distance contributions to the
%\,$\Sigma^{+}\to p\mu^{+}\mu^{-}$\, rate are well below the
%observed value.
In Sec.~\ref{ld} we study the long-distance contributions mediated
by a real or a virtual photon. These can be parameterized by four
(complex) gauge-invariant form-factors~\cite{Bergstrom:1987wr}. We
determine the imaginary parts of these form factors from unitarity.
The real parts of two of the form factors can be reasonably assumed
to be constant as a first approximation and can then be extracted from
the measured rate and asymmetry parameter for \,$\Sigma^{+}\to p\gamma$\,
up to a fourfold ambiguity. The real parts of the two remaining
form-factors cannot be extracted from experiment at present, and so we
estimate them using vector-meson-dominance models.
Finally, in Sec.~\ref{sum} we combine all these results to present
the predictions for the rates and spectra of the two modes
$\Sigma^{+}\to p\mu^{+}\mu^{-},\,p e^+e^-$.\,
Before concluding, we discuss the implications of our analysis for
the possibility that new physics could be present in the recent
measurement by HyperCP.

\section{Short-distance contributions\label{sd}}

The short-distance effective Hamiltonian responsible for
\,$\Sigma^+\to p \ell^+\ell^-$\,  contains contributions originating from the
$Z$-penguin, box, and electromagnetic-penguin diagrams. It is given
by~\cite{Shifman:1976de,Buchalla:1995vs}
\begin{eqnarray}
{\cal H}_{\rm eff}^{} &=& \frac{G_F^{}}{\sqrt{2}} V_{ud}^* V_{us}^{}\,
\bigl[ \bigl(z_{7V}^{}+\tau y_{7V}^{}\bigr) O_{7V}^{}+ \tau y_{7A}^{} O_{7A}^{}\bigr]
+ \frac{G_F^{}}{\sqrt{2}} \sum_j V_{jd}^* V_{js}^{}\, c^j_{7\gamma}O_{7\gamma}  \,\,,
\end{eqnarray}
where  $V_{kl}^{}$  are the elements of the Cabibbo-Kobayashi-Maskawa (CKM) matrix~\cite{ckm},
$z$, $y$, and $c$ are the Wilson coefficients,
$\,\tau =- V_{td}^*V_{ts}^{}/\bigl(V_{ud}^*V_{us}^{}\bigr)$,\,  and
\begin{eqnarray}
O_{7V}^{}  &=&  \bar d\gamma^\mu(1-\gamma_5^{})s\, \bar\ell^-\gamma_\mu^{}\ell^+  \,\,,
\hspace{2em}
O_{7A}^{}  \,\,=\,\,
\bar d\gamma^\mu(1-\gamma_5^{})s\, \bar\ell^-\gamma_\mu^{}\gamma_5^{}\ell^+  \,\,,
\nonumber\\
O_{7\gamma}^{} &=& \frac{e}{16\pi^2}\, \bar d \sigma^{\mu\nu} F_{\mu\nu}^{}
\bigl[ m_s^{} (1+\gamma_5^{}) + m_d^{} (1-\gamma_5^{})\bigr] s   \,\,,
\end{eqnarray}
with  $F_{\mu\nu}^{}$  being the photon field-strength tensor.
The contribution of  $O_{7\gamma}^{}$  to  \,$\Sigma^+\to p \ell^+\ell^-$\,  occurs
via the photon converting to a lepton pair.  The total short-distance contribution
to the  \,$\Sigma^+\to p\ell^+\ell^-$\,  amplitude is then given by
\begin{eqnarray}
&& \hspace*{-4ex}
{\cal M}(\Sigma^+\to p\ell^+\ell^-) \,\,=\,\,
\bigl\langle p\ell^+\ell^-\bigr|{\cal H}_{\rm eff}^{}\bigl|\Sigma^+\bigr\rangle
\\
&=&  \frac{G_F^{}}{\sqrt{2}}
\left\{ V_{ud}^*V_{us}^{} \bigl[ (z_{7V}^{} + \tau y_{7V}^{})
\langle p|\bar d\gamma^\mu(1-\gamma_5^{})s|\Sigma^+\rangle \bar\ell^-\gamma_\mu^{}\ell^+
+ \tau y_{7A}^{} \langle p|\bar d\gamma^\mu(1-\gamma_5^{})s|\Sigma^+\rangle
\bar\ell^-\gamma_\mu^{}\gamma_5^{}\ell^+ \bigr]
\vphantom{\sum_i}  \right.
\nonumber\\
&&-\, \left. \sum_j
V_{jd}^*V_{js}^{}\,\frac{i\alpha\,c^j_{7\gamma}}{2\pi q^2} \bigl[ (m_s^{} + m_d^{})
\langle p|\bar d \sigma^{\mu\nu}q_\nu^{} s|\Sigma^+\rangle
+ (m_s^{}-m_d^{})\langle p|\bar d\sigma^{\mu\nu}q_\nu^{}\gamma_5^{}s|\Sigma^+\rangle \bigr]\,
\bar\ell^-\gamma_\mu^{}\ell^+ \right\}   \,\,, \nonumber
\end{eqnarray}
where  \,$q=p_\Sigma^{}-p_p^{}$.

To obtain the corresponding branching ratio, one needs to know the hadronic matrix elements.
Employing the leading-order strong Lagrangian in chiral perturbation theory ($\chi$PT),
given in Eq.~(\ref{Ls1}), we find
\begin{eqnarray}
\langle p|\bar d \gamma^\mu s|\Sigma^+\rangle  \,\,=\,\, -\bar p\gamma^\mu \Sigma  \,\,,
\,\,&&\,\,
\langle p|\bar d\gamma^\mu\gamma_5^{} s|\Sigma^+\rangle  \,\,=\,\,
(D-F)\, \bar p\gamma^\mu\gamma_5^{} \Sigma   \,\,,
\end{eqnarray}
where  \,$D=0.80$\, and  \,$F=0.46$\,  from fitting to hyperon
semileptonic decays, and  using quark-model
results~\cite{Donoghue:1992dd} we obtain
\begin{eqnarray}
\langle p|\bar d\sigma^{\mu\nu}s|\Sigma^+\rangle  \,\,=\,\,
c_\sigma^{}\,\bar p\sigma^{\mu\nu}\Sigma   \,\,,
\,\,&&\,\,
\langle p|\bar d\sigma^{\mu\nu}\gamma_5^{} s|\Sigma^+\rangle  \,\,=\,\,
c_\sigma^{}\, \bar p\sigma^{\mu\nu} \gamma_5^{} \Sigma   \,\,,
\end{eqnarray}
where  \,$c_\sigma^{}=-1/3$.\,
Furthermore, we adopt the CKM-matrix elements given in Ref.~\cite{pdg}, the typical Wilson
coefficients obtained in the literature~\cite{Shifman:1976de,Buchalla:1995vs}, namely
\,$z_{7V}^{} = -0.046\alpha$,\,  \,$y_{7V}^{} = 0.735\alpha$,\,
\,$y_{7A}^{} = -0.700\alpha$\,~\cite{Buchalla:1995vs}, and  $c_{7\gamma}^j$ being
dominated by  \,$c^c_{7\gamma} =0.13$\,~\cite{Shifman:1976de}, and the quark masses
\,$m_d^{}=9\rm\,MeV$\,  and  \,$m_s^{}=120\rm\,MeV$.\,

The resulting branching ratio for  \,$\Sigma^+\to p\mu^+\mu^-$\, is about
\,$10^{-12}$,\,  which is way below the observed value.  There are uncertainties
in the hadronic matrix elements, the Wilson
coefficients, and the CKM-matrix elements, but these uncertainties will not change
this result by orders of magnitude.  We therefore conclude that in the SM
the short-distance contribution is too small to explain the HyperCP data on
\,$\Sigma^+\to p\mu^+\mu^-$.

Now, a large branching ratio for  \,$\Sigma^+\to p \ell^+\ell^-$\,  may be related to
the large observed branching ratio for  \,$\Sigma^+\to p\gamma$,\,  compared with
their respective short-distance contributions. With only the short-distance
contribution to  \,$\Sigma^+\to p \gamma$\,  within the SM,  the branching ratio is
predicted to be much smaller than the experimental value~\cite{He:1999ik}.
However, beyond the SM it is possible to have an enhanced short-distance
contribution to \,$\Sigma^+\to p \gamma$\,~\cite{He:1999ik}  which would enhance
the amplitude for  \,$\Sigma^+\to p\mu^+\mu^-$.\,  The origin of the enhancement
may be from new interactions such as  $W_L$-$W_R$ mixing in left-right symmetric
models and left-right squark mixing in supersymmetric models~\cite{He:1999ik}.
These types of interactions have small effects on other related flavor-changing
processes such as $K^0$-$\bar K^0$ mixing, but can have large effects  on
\,$\Sigma^+\to p\gamma$\,  and therefore also on  \,$\Sigma^+\to p\ell^+\ell^-$.\,
Thus the observed branching ratio for  \,$\Sigma^+\to p\gamma$\,  can be reproduced
even if one assumes that there is only the short-distance contribution.
More likely, however, the enhancement is due to long-distance contributions within
the SM. In the next section we present the most complete estimate
possible at present for these long-distance contributions.

\section{Long-distance contributions\label{ld}}

In this section we deal with the contributions to
$\,\Sigma^{+}\to p\ell^{+}\ell^{-}\,$  that are mediated by a~photon. For a real
intermediate photon there are two form factors that can be extracted
from the weak radiative hyperon decay  $\,B_{i}^{}\to B_{f}^{}\gamma\,$ and
are usually parameterized by the effective Lagrangian
\begin{equation}
{\cal L} \,\,=\,\,
\frac{eG_{F}}{2}\, \bar{B}_{f}\left(a+b\gamma_{5}\right)\sigma^{\mu\nu}B_{i}\, F_{\mu\nu} \,\,.
\label{radff}
\end{equation}
The two form factors, $a$ and $b$, are related to the width and decay distribution of
the radiative decay by
\begin{eqnarray}
\Gamma(B_{i}\to B_{f}\gamma) &=&
\frac{G_{F}^{2}e^{2}}{\pi}\left(|a|^{2}+|b|^{2}\right)\omega^{3} \,\,,
\end{eqnarray}
\begin{eqnarray}
\frac{d\Gamma}{d\cos\theta}  \,\,\sim\,\,  1+\alpha\,\cos\theta   \,\,,   \hspace{2em}
\alpha  \,\,=\,\,  \frac{2\,Re\,(ab^{*})}{|a|^{2}+|b|^{2}} \,\,,
\end{eqnarray}
where $\omega$ is the photon energy, and  $\theta$ is the angle between the spin of
$B_i^{}$ and the three-momentum of  $B_f^{}$.
The measured values for  $\,\Sigma^{+}\to p\gamma\,$  are~\cite{pdg}
\begin{eqnarray}   \label{Spgdata}
\Gamma(\Sigma^{+}\to p\gamma) \,\,=\,\, (10.1\pm 0.4)\times10^{-15}{\rm~MeV}  \,\,,
\hspace{2em}   \alpha \,\,=\,\, -0.76 \pm 0.08  \,\,.
\end{eqnarray}

When the photon is a virtual one, there are two additional
form-factors, and the total amplitude can be parameterized as
\begin{eqnarray}   \label{M_BBg}
{\cal M}(B_i\to B_f\gamma^*) &=&
- e G_{F}^{}\, \bar{B}_f^{} \left[ i\sigma^{\mu\nu}q_\mu^{}(a+b\gamma_5^{})
+(q^2\gamma^\nu-q^\nu\!\!\not{\!q}) (c+d\gamma_5^{}) \right] B_i^{}\, \varepsilon_\nu^{*}   \,\,,
\end{eqnarray}
where  $q$  is the photon four-momentum.
We note that the $a$ and $c$ ($b$ and $d$) terms are parity conserving (violating).
The corresponding amplitude for  $\,B_{i}^{}\to B_{f}^{}\ell^{+}\ell^{-}\,$  is then
\begin{eqnarray}
{\cal M}(B_i\to B_f\ell^+\ell^-) &=&
\frac{-i e^2 G_{F}^{}}{q^2}\,
\bar{B}_{f}^{}\left(a+b\gamma_5^{}\right)\sigma_{\mu\nu}^{}q^{\mu}B_{i}^{}\,
\bar\ell^{-}\gamma^\nu\ell^{+}
\nonumber \\ &&   \vphantom{\int^|}
-\,\, e^2 G_{F}^{}\, \bar{B}_{f}^{}\gamma_\mu^{}(c+d\gamma_{5}^{})B_{i}^{}\,
\bar\ell^{-}\gamma^{\mu}\ell^{+}   \,\,,  \label{ffabcd}
\end{eqnarray}
where now  $\,q=p_{\ell^+}^{}+p_{\ell^-}^{}.\,$
In general  $a$, $b$, $c$, and $d$  depend on~$q^2$, and  for  $\,\Sigma^{+}\to p\gamma^*\,$
the first two are constrained at  $\,q^2=0\,$  by the data in Eq.~(\ref{Spgdata}) as
\begin{eqnarray}  \label{Spgcons}
|a(0)|^{2}+|b(0)|^{2} &=& (15.0\pm 0.3)^{2}{\rm ~MeV}^{2}   \,\,,
\nonumber \\
{\rm Re}\,\bigl(a(0)\,b^*(0)\bigr) &=& (-85.3\pm 9.6) {\rm ~MeV}^{2}   \,\,.
\end{eqnarray}
These form factors are related to the ones in Ref.~\cite{Bergstrom:1987wr} by
\begin{eqnarray}
a  \,\,=\,\,  2i b_1^{}  \,\,,  \hspace{2em}
b  \,\,=\,\,  2i b_2^{}  \,\,,  \hspace{2em}
c  \,\,=\,\,  \frac{i a_1^{}}{q^2}  \,\,,  \hspace{2em}
d  \,\,=\,\,  -\frac{i a_2^{}}{q^2}  \,\,.
\end{eqnarray}

As we will estimate later on, these form factors have fairly mild $q^2$-dependence.
If they are taken to be constant, by integrating numerically over phase space we can
determine the branching ratios of  \,$\Sigma^{+}\to p\ell^{+}\ell^{-}$\, to be,
with $a$ and $b$ in MeV,
\begin{subequations}  \label{rateres}
\begin{eqnarray}
{\cal B}(\Sigma^{+}\to p \mu^{+}\mu^{-}) &=& \left[ 2.00 \left(|a|^{2}+|b|^{2}\right)
-1.60 \left(|a|^{2}-|b|^{2}\right) \right] \times 10^{-10} \nonumber \\
&&+\,\, \left( 1.05\, |c|^{2}+ 18.2\, |d|^{2}\right)\times 10^{-6}\nonumber \\
&&+\,\, \left[ 0.29 {\rm~Re}\,(ac^{*}) - 16.1 {\rm~Re}\,(bd^{*})\right]
\times 10^{-8}  \,\,,
\end{eqnarray}
\begin{eqnarray}
{\cal B}(\Sigma^{+}\to p e^{+}e^{-}) &=& \left[ 4.22 \left(|a|^{2}+|b|^{2}\right)
-0.21 \left(|a|^{2}-|b|^{2}\right) \right] \times 10^{-8} \nonumber \\
&&+\,\, \left( 5.38\, |c|^{2}+ 15.9\, |d|^{2}\right)\times 10^{-5}\nonumber \\
&&+\,\, \left[ 1.51 {\rm~Re}\,(ac^{*}) - 21.1 {\rm~Re}\,(bd^{*})\right]
\times 10^{-7}   \,\,.
\end{eqnarray}
\end{subequations}
If the form factors have  $q^2$-dependence, the expression is different, and
the rate should be calculated with the formula which we give in Appendix~\ref{diffrate}.

\subsection{Imaginary parts of the form factors from unitarity}

The form factors which contribute to the weak radiative hyperon decays have been studied
in chiral perturbation theory~\cite{Neufeld:1992hb,Jenkins:1992ab,Bos:1996ig}.
The imaginary parts of  $a$ and $b$   for  $\,\Sigma^{+}\to p\gamma$\,
have been determined from unitarity with different results in the literature.
Neufeld~\cite{Neufeld:1992hb} employed relativistic baryon $\chi$PT to find, for $\,q^2=0$,
\begin{eqnarray}   \label{neufeld}
{\rm Im}\, a(0) \,\,=\,\,  2.60~{\rm MeV}   \,\,,  \hspace{2em}
{\rm Im}\, b(0) \,\,=\,\, -1.46~{\rm MeV}
\end{eqnarray}
in the notation of Eq.~(\ref{radff}), whereas Jenkins {\it et al.}~\cite{Jenkins:1992ab}
using the heavy-baryon formulation  obtained
\begin{eqnarray}   \label{jenkins}
{\rm Im}\, a(0) \,\,=\,\,  6.18~{\rm MeV}   \,\,,  \hspace{2em}
{\rm Im}\, b(0) \,\,=\,\, -0.53~{\rm MeV}   \,\,.
\end{eqnarray}
Because of this disagreement, and since we also need the imaginary parts of the form factors
$c$ and $d$,  we repeat here the unitarity calculation employing both the relativistic
and heavy baryon approaches.

\begin{figure}[htb]
\includegraphics{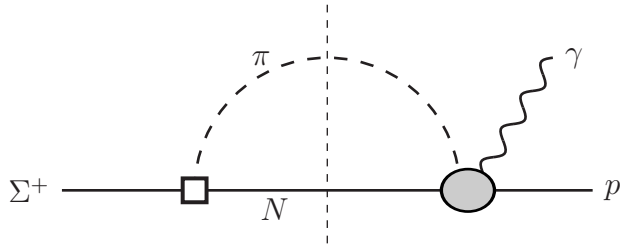}
\caption{Unitarity cut.\label{fig_cut}}
\end{figure}

Our strategy to derive the imaginary parts of the four form-factors
in Eq.~(\ref{ffabcd}) from unitarity is illustrated in
Fig.~\ref{fig_cut}. As the figure shows, these imaginary parts
can be determined from the amplitudes for the weak nonleptonic
decays $\,\Sigma^{+}\to p\pi^{0}\,$  and  $\,\Sigma^{+}\to n\pi^{+}$\,
(the vertex indicated by a square in Fig.~\ref{fig_cut})  as well as
the reactions $\,N\pi\to N\gamma^*$\, (the vertex indicated by a blob in
Fig.~\ref{fig_cut}).
The weak decays have been measured~\cite{pdg}, and we express their
amplitudes as\footnote{ We have taken the nonzero
elements of $\gamma_5^{}$ to be positive.}
\begin{eqnarray}   \label{M_SNpi}
{\cal M}(\Sigma^{+}\to N\pi) &=& i G_{F}^{}m_{\pi^{+}}^{2}\, \bar{N}
\left(A_{N\pi} - B_{N\pi}\gamma_{5}^{}\right) \Sigma  \,\,,
\end{eqnarray}
where
\begin{eqnarray}   \label{ABnlhd}
A_{n\pi^+}  \,\,=\,\,   0.06  \,\,,  \,\,\, &&
B_{n\pi^+}  \,\,=\,\,  18.53  \,\,,  \nonumber \\
A_{p\pi^0}  \,\,=\,\,  -1.43  \,\,,  &&
B_{p\pi^0}  \,\,=\,\,  11.74  \,\,.
\end{eqnarray}
Following Refs.~\cite{Neufeld:1992hb,Jenkins:1992ab}, we adopt the  $\,N\pi\to p\gamma^*\,$
amplitudes derived in lowest-order  $\chi$PT.

\begin{figure}[htb]
\includegraphics{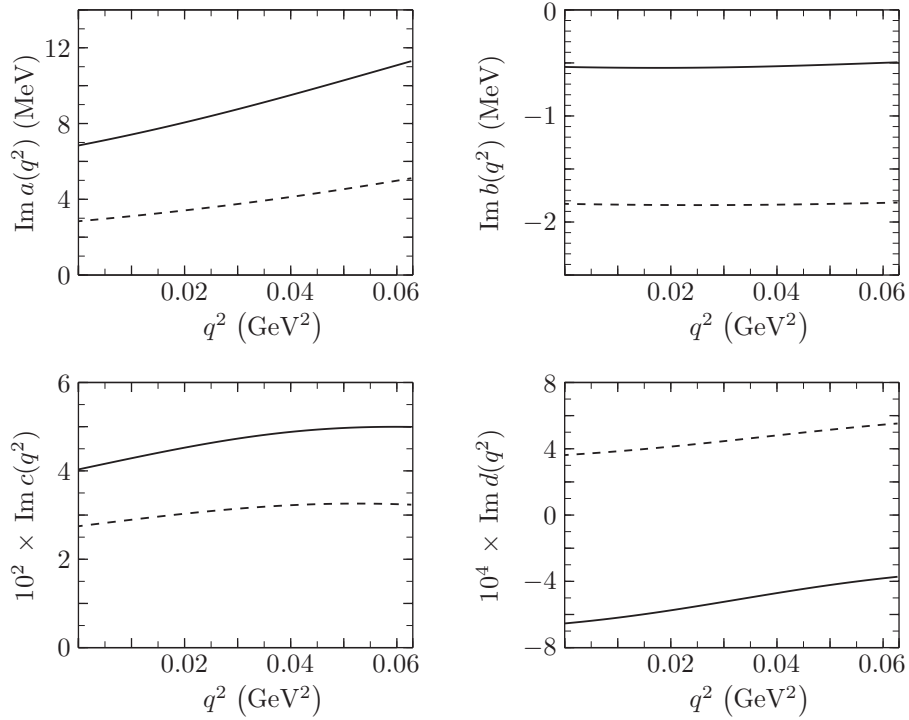}
\caption{Imaginary parts of the form factors in  \,$\Sigma^+\to p\gamma^*$,\,  obtained
using heavy baryon $\chi$PT  (solid lines) and relativistic baryon $\chi$PT (dashed lines).
\label{fig_imF}}
\end{figure}

We present the details of our unitarity calculation in  Appendix~\ref{imabcd}.
The results in the relativistic and heavy baryon approaches are given in Eqs.~(\ref{imFF_r})
and~(\ref{imFF_hb}), respectively.  In Fig.~\ref{fig_imF} we display the two sets of form
factors for  $\,0\le q^2\le(m_\Sigma^{}-m_N^{})^2$.\,
We note that, although only the  $\,\Sigma^+\to n\pi^+\,$  transition contributes to
the heavy-baryon form-factors at leading order, the sizable difference between the
${\rm Im}\,a$, or  ${\rm Im}\,c$,  curves arises mainly from relativistic corrections,
which reduce the heavy-baryon numbers by about 50\%.
On the other hand, the difference between the  ${\rm Im}\,b$, or  ${\rm Im}\,d$,  curves
is due not only to relativistic corrections, but also to  $A_{n\pi^+}$  being much smaller
than  $A_{p\pi^0}$.

To compare with the numbers in Eqs.~(\ref{neufeld}) and~(\ref{jenkins}) calculated
in earlier work,  we find from the relativistic formulas in Eq.~(\ref{imFF_r})
\begin{eqnarray}   \label{imab_r}
{\rm Im}\, a(0) \,\,=\,\,  2.84~{\rm MeV}   \,\,, \hspace{2em}
{\rm Im}\, b(0) \,\,=\,\,  -1.83~{\rm MeV}   \,\,,
\end{eqnarray}
and from the heavy-baryon results in Eq.~(\ref{imFF_hb})
\begin{eqnarray}   \label{imab_hb}
{\rm Im}\, a(0) \,\,=\,\,  6.84~{\rm MeV}   \,\,,  \hspace{2em}
{\rm Im}\, b(0) \,\,=\,\,  -0.54~{\rm MeV}   \,\,.
\end{eqnarray}
Thus our relativistic results are close to those in Eq.~(\ref{neufeld}), from
Ref.~\cite{Neufeld:1992hb}, and our heavy-baryon numbers to those in
Eq.~(\ref{jenkins}),  from Ref.~\cite{Jenkins:1992ab}.\footnote{
Our heavy-baryon expressions for  ${\rm Im}\, a(0)$  and  ${\rm Im}\, b(0)$ are
identical to those in Ref.~\cite{Jenkins:1992ab}, except that
their ${\rm Im}\, a(0)$  formula has one of the overall factors of
\,$1/(m_\Sigma^{}-m_N^{})$\, apparently coming from their approximating
$\bigl[(m_\Sigma^{}-m_N^{})^2-m_\pi^2\bigr]{}^{1/2}$  as  $m_\Sigma^{}-m_N^{}$.
This is the main reason for the value of ${\rm Im}\, a(0)$ in Eq.~(\ref{jenkins})
being smaller than that in Eq.~(\ref{imab_hb}).
}
These two sets of numbers are different for the reasons mentioned in the preceding
paragraph.

\subsection{Real parts of the form factors}

The real parts of the form factors cannot be completely predicted at present from
experimental input alone. For  ${\rm Re}\,a(q^{2})$ and ${\rm Re}\,b(q^{2})$,
the values at $\,q^{2}=0\,$ can be extracted from  Eq.~(\ref{Spgcons}) after using
Eq.~(\ref{imab_r}) or~(\ref{imab_hb})  for the imaginary parts.
Thus the relativistic numbers in Eq.~(\ref{imab_r}) lead to the four sets of solutions
\begin{eqnarray}   \label{reab_r}
{\rm Re}\, a(0) \,\,=\,\, \pm 13.3{\rm~MeV} \,\,,   \,\,&&\,\,
{\rm Re}\, b(0) \,\,=\,\, \mp  6.0{\rm~MeV}  \,\,,
\nonumber \\
{\rm Re}\, a(0) \,\,=\,\, \pm  6.0{\rm~MeV}  \,\,,   \,\,&&\,\,
{\rm Re}\, b(0) \,\,=\,\, \mp 13.3{\rm~MeV} \,\,,
\end{eqnarray}
while the heavy-baryon results in Eq.~(\ref{imab_hb}) imply
\begin{eqnarray}    \label{reab_hb}
{\rm Re}\, a(0) \,\,=\,\, \pm 11.1{\rm~MeV} \,\,,   \,\,&&\,\,
{\rm Re}\, b(0) \,\,=\,\, \mp  7.3{\rm~MeV}  \,\,,
\nonumber \\
{\rm Re}\, a(0) \,\,=\,\, \pm  7.3{\rm~MeV}  \,\,,  \,\,&&\,\,
{\rm Re}\, b(0) \,\,=\,\, \mp 11.1{\rm~MeV} \,\,.
\end{eqnarray}
Since these numbers still cannot be predicted reliably within the
framework of $\chi$PT~\cite{Neufeld:1992hb,Jenkins:1992ab}, we
will assume that
\begin{eqnarray}   \label{reab}
{\rm Re}\, a(q^2)  \,\,=\,\,  {\rm Re}\, a(0)  \,\,,  \,\,&&\,\,
{\rm Re}\, b(q^2)  \,\,=\,\,  {\rm Re}\, b(0)  \,\,,
\end{eqnarray}
where the  $\,q^2=0\,$  values are those in Eqs.~(\ref{reab_r})
and~(\ref{reab_hb}) in the respective approaches. This assumption
is also reasonable in view of the fairly mild $q^2$-dependence of
the imaginary parts seen in Fig.~\ref{fig_imF}, and of the real
parts of $c$ and $d$ below.
In predicting the  \,$\Sigma^+\to p\ell^+\ell^-$\,  rates in the
following section,  we will use the 8 sets of possible solutions in
Eqs.~(\ref{reab_r}) and~(\ref{reab_hb}).

\begin{figure}[htb]
\includegraphics{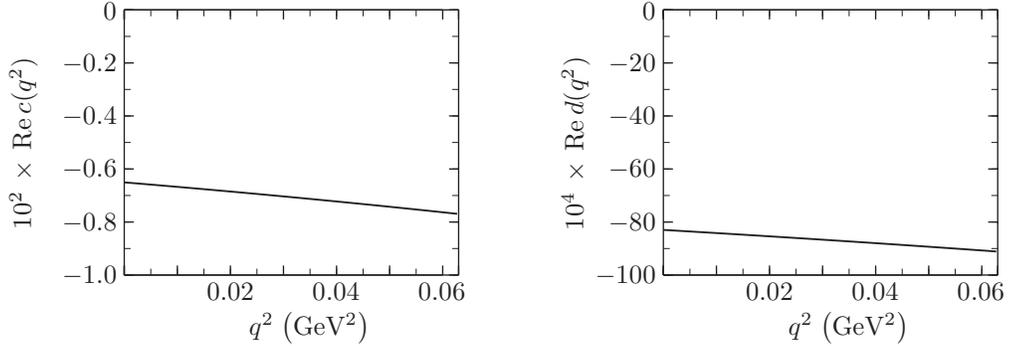}
\caption{Real parts of $c$ and $d$.\label{fig_reF}}
\end{figure}

The real parts of $c$ and $d$ cannot be extracted from experiment at present.
Our interest here, however, is in predicting the SM contribution, and therefore we
need to estimate them.  To do so, we employ a~vector-meson-dominance assumption,
presenting the details in  Appendix~\ref{recd}.  The results for  ${\rm Re}\,c(q^2)$
and  ${\rm Re}\,d(q^2)$  are given in Eqs.~(\ref{rec}) and~(\ref{red}), respectively.
In Fig.~\ref{fig_reF} we display the two form factors for
$\,0\le q^2\le(m_\Sigma^{}-m_N^{})^2$.\,
We can see from Figs.~\ref{fig_imF} and~\ref{fig_reF} that $c$ is dominated by
its imaginary part,  but that  $d$ is mostly real.

\section{Results and conclusions\label{sum}}

We can now evaluate the rates and spectra of  \,$\Sigma^+\to p\ell^+\ell^-$\, resulting from
the various standard-model contributions. Since the short-distance contributions
discussed in  Sec.~\ref{sd} are very small, we shall neglect them. Consequently,
the rates are determined  by   the various form factors in
\,$\Sigma^+\to p\gamma^*$\,  calculated in the preceding section
and applied in Eq.~(\ref{diffrateform}).

In Table~\ref{rates}, we have collected the branching ratios of
\,$\Sigma^+\to p\mu^+\mu^-$\,  and  \,$\Sigma^+\to p e^+e^-$\,  corresponding to the 8~sets
of solutions in  Eqs.~(\ref{reab_r}) and~(\ref{reab_hb}), under the assumption of
Eq.~(\ref{reab}) for  ${\rm Re}\,a$ and~${\rm Re}\,b$.  The real parts of $c$ and $d$ in
Eqs.~(\ref{rec}) and~(\ref{red}) are used in all the unbracketed branching ratios.
For the imaginary parts of the form factors, the expressions in Eq.~(\ref{imFF_r})
[Eq.~(\ref{imFF_hb})] contribute to the unbracketed branching ratios in the upper (lower)
half of this table.
Within each pair of square brackets, the first number is the branching ratio
obtained without contributions from both $c$ and $d$, whereas the second number
is the branching ratio calculated with only the real parts of all the form factors.

\begin{table}[ht]
\caption{\label{rates}%
Branching ratios of  \,$\Sigma^+\to p\mu^+\mu^-, p e^+e^-$\, in the standard model.
The unbracketed branching ratios receive contributions from all the form factors,
with the expressions in  Eq.~(\ref{imFF_r}) [Eq.~(\ref{imFF_hb})]  for the imaginary
parts contributing to the numbers in the first (last) four rows.
Within each pair of square brackets, the first number has been obtained with
$\,c=d=0$, and the second with only the real parts of all the form factors.}
\centering
\footnotesize \vskip 0.5\baselineskip
\begin{tabular}{@{\hspace{1em}}r@{\hspace{1em}}|@{\hspace{1em}}r@{\hspace{1em}}|@{\hspace{1em}}c
@{\hspace{1em}}|@{\hspace{1em}}r@{\hspace{1em}}}
\hline \hline
${\rm Re}\,a$ (MeV) $\displaystyle\vphantom{\int}$  &  ${\rm Re}\,b$ (MeV) &
$10^8\,{\cal B}\bigl(\Sigma^+\to p\mu^+\mu^-\bigr)$  &
$10^6\,{\cal B}\bigl(\Sigma^+\to p e^+e^-\bigr)$
\\ \hline
   13.3 \hspace{3ex} &  $-$6.0 \hspace{3ex} & 1.6 \,\, [2.2,\,\,1.3] &  9.1 \,\, [9.2,\,\,8.6] \hspace{2ex} \\
$-$13.3 \hspace{3ex} &     6.0 \hspace{3ex} & 3.4 \,\, [2.2,\,\,3.1] &  9.4 \,\, [9.2,\,\,8.8] \hspace{2ex} \\
    6.0 \hspace{3ex} & $-$13.3 \hspace{3ex} & 5.1 \,\, [6.7,\,\,4.7] &  9.6 \,\, [9.8,\,\,9.0] \hspace{2ex} \\
 $-$6.0 \hspace{3ex} &    13.3 \hspace{3ex} & 9.0 \,\, [6.7,\,\,8.6] & 10.1 \,\, [9.8,\,\,9.5] \hspace{2ex} \\
\hline
   11.1 \hspace{3ex} &  $-$7.3 \hspace{3ex} & 2.3 \,\, [2.9,\,\,1.5] &  9.3 \,\, [9.3,\,\,7.2] \hspace{2ex} \\
$-$11.1 \hspace{3ex} &     7.3 \hspace{3ex} & 4.5 \,\, [2.9,\,\,3.7] &  9.6 \,\, [9.3,\,\,7.5] \hspace{2ex} \\
    7.3 \hspace{3ex} & $-$11.1 \hspace{3ex} & 4.0 \,\, [5.1,\,\,3.2] &  9.5 \,\, [9.6,\,\,7.4] \hspace{2ex} \\
 $-$7.3 \hspace{3ex} &    11.1 \hspace{3ex} & 7.3 \,\, [5.1,\,\,6.4] & 10.0 \,\, [9.6,\,\,7.8] \hspace{2ex} \\
\hline \hline
\end{tabular}
\bigskip
\end{table}

\begin{figure}[tb]
\includegraphics{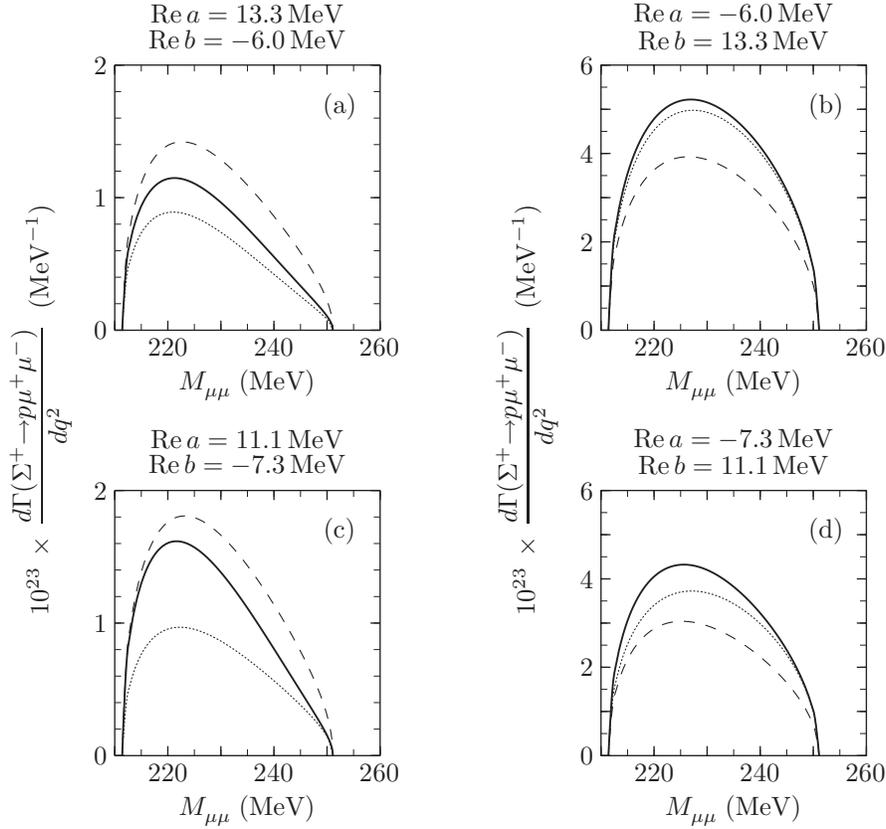}
\caption{Invariant-mass distributions of the lepton pair in
\,$\Sigma^+\to p\mu^+\mu^-$\,  corresponding to
the smallest and largest branching ratios for the (a,b) relativistic and
(c,d) heavy baryon cases in Table~\ref{rates}.
In all distribution figures, each solid curve receives contributions from
all the form factors, each dashed curve has been obtained with
\,$c=d=0$,\, and each dotted curve involves no imaginary parts of all the form factors.
\label{fig_BRmu}}
\end{figure}

\begin{figure}[htb]
\includegraphics{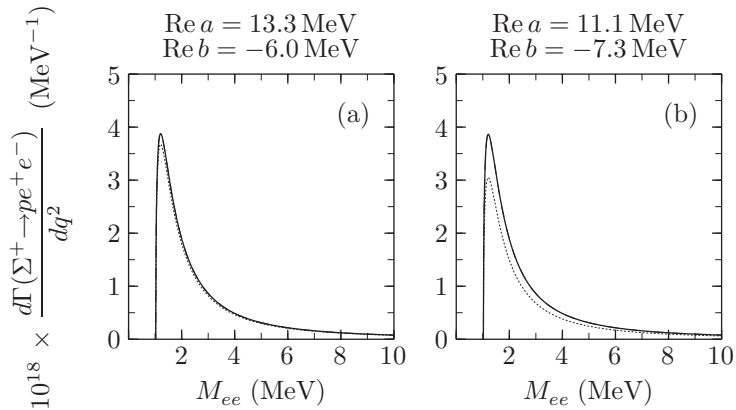}
\caption{Low-mass portion of the invariant-mass distributions of the lepton pair in
\,$\Sigma^+\to p e^+e^-$\,  corresponding to two of the branching ratios in
Table~\ref{rates},  for the (a) relativistic and (b) heavy baryon cases.
\label{fig_BRee}}
\end{figure}

In Fig.~\ref{fig_BRmu} we show the invariant-mass distributions of the $\mu^+\mu^-$
pair, with  \,$M_{\mu\mu}^{}=\sqrt{q^2}$,\,  that correspond to the smallest and largest
rates of \,$\Sigma^+\to p\mu^+\mu^-$\,  listed in Table~\ref{rates}
for both the relativistic baryon [(a)\,and\,(b)] and heavy baryon [(c)\,and\,(d)] cases.
For \,$\Sigma^+\to p e^+e^-$,\,  the mass distributions of the  $e^+e^-$ pair,
two of which are displayed in Fig.~\ref{fig_BRee}, differ very little from each other
and are strongly peaked at low  \,$M_{ee}^{}=\sqrt{q^2}$.\,
Also shown in the figures are the distributions obtained with  \,$c=d=0$\,  (dashed curves),
as well as those without contributions from the imaginary parts of all the form
factors (dotted curves).

We can see from Table~\ref{rates}, Fig.~\ref{fig_BRmu}, and Fig.~\ref{fig_BRee} that
the effect of the $c$ and $d$ contributions on the total rates can be up to nearly~40\%
in  \,$\Sigma^+\to p\mu^+\mu^-$,\,  but it is much smaller in  \,$\Sigma^+\to p e^+e^-$.\,
Furthermore, the contributions of the imaginary parts of the form factors can be
as large as~35\%  to the  $p\mu^+\mu^-$  rate and roughly~20\%  to the  $pe^+e^-$  rate.
This implies that a careful analysis of experimental results, especially in
the case of  \,$\Sigma^+\to p\mu^+\mu^-$,\,  should take into account the imaginary
parts of the form factors.

For  \,$\Sigma^+\to p\mu^+\mu^-$,\,  HyperCP measured the branching ratio to be
\,$\bigl(8.6_{-5.4}^{+6.6}\pm5.5\bigr)\times10^{-8}$\,~\cite{Park:2005ek}.
It is evident that all the predictions in  Table~\ref{rates}  for
the  \,$p\mu^+\mu^-$\,  mode corresponding to the different sets of form factors
fall within the experimental range.
For  \,$\Sigma^+\to p e^+e^-$,\,  the branching ratio can be inferred from
the experimental results given in  Ref.~\cite{ang}, which reported the width ratio
\,$\Gamma(\Sigma^+\to p e^+e^-)/\Gamma(\Sigma^+\to p\pi^0)=(1.5\pm0.9)\times10^{-5}$\,
and interpreted the observed events as proceeding from
\,$\Sigma^+\to p\gamma^*$,\,  based on the very low invariant-masses of
the $e^+e^-$ pair.\footnote{We note that the upper limit of
\,$7\times10^{-6}$\,  quoted in Ref.~\cite{pdg} and obtained in Ref.~\cite{ang}
is for the presence of weak neutral currents in \,$\Sigma^+\to p e^+e^-$\,  and
not for the branching ratio of this mode.}
This number, in conjunction with the current data on  \,$\Sigma^+\to p\pi^0$\,~\cite{pdg},
translates into  \,${\cal B}(\Sigma^+\to p e^+e^-)=(7.7\pm4.6)\times10^{-6}$.\,
Clearly, the results for  the  \,$p e^+e^-$\,  mode in Table~\ref{rates} are well
within the experimentally allowed range.
Based on the numbers in Table~\ref{rates}, we may then conclude that within
the standard model
\begin{eqnarray}   \label{results}
\begin{array}{c}   \displaystyle
1.6\times10^{-8}  \,\,\le\,\,  {\cal B}\bigl(\Sigma^+\to p\mu^+\mu^-\bigr)
\,\,\le\,\,  9.0\times10^{-8}   \,\,,
\vspace{2ex} \\   \displaystyle
9.1\times10^{-6}  \,\,\le\,\,  {\cal B}\bigl(\Sigma^+\to p e^+e^-\bigr)
\,\,\le\,\,  10.1\times10^{-6}   \,\,.
\end{array}
\end{eqnarray}

The agreement above between the predicted and observed rates of
\,$\Sigma^+\to p\ell^+\ell^-$\,  indicates that these decays are dominated
by long-distance contributions.
However, the predicted range for  ${\cal B}(\Sigma^+\to p\mu^+\mu^-)$
is sufficiently wide that we cannot rule out the possibility of a new-physics
contribution of the type suggested by HyperCP~\cite{Park:2005ek}.
Motivated by the narrow distribution of dimuon masses of the events
they observed, they proposed that the decay could proceed via a new intermediate
particle of mass  $\sim$\,214\,MeV,\, with a branching ratio of
\,$\bigl(3.1_{-1.9}^{+2.4}\pm1.5\bigr)\times10^{-8}$\,~\cite{Park:2005ek}.
For this hypothesis to be realized, however, the new physics would have to
dominate the decay.
It will be interesting to see if this hypothesis will be confirmed by
future measurements.

Finally, we observe that the smaller numbers
\,${\cal B}(\Sigma^+\to p\mu^+\mu^-)\sim2\times10^{-8}$\,  in Table~\ref{rates}
correspond to the mass distributions peaking at lower masses,
\,$M_{\mu\mu}^{}\sim220\rm\,MeV$,\,  in Fig.~\ref{fig_BRmu}.
It is perhaps not coincidental that these numbers are similar to
the branching ratio and new-particle mass, respectively, in the HyperCP
hypothesis above.
This may be another indication that it is not necessary to invoke new
physics to explain the HyperCP results.

\begin{acknowledgments}
We thank HyangKyu Park for conversations.
The work of X.G.H. was supported in part by the National Science Council under NSC grants.
The work of G.V. was supported in part by DOE under contract
number DE-FG02-01ER41155.
\end{acknowledgments}

\appendix

\section{Differential rate of $\bm{\Sigma^+\to p\ell^+\ell^-}$\label{diffrate}}

If the form factors have $q^2$-dependence, before integrating over
phase space to obtain the branching ratio we should use
\begin{eqnarray}   \label{diffrateform}
&& \hspace*{-3em} \frac{d\Gamma(\Sigma^+\to p\ell^+\ell^-)}{d
q^2\, dt}  \,\,=\,\, \frac{\alpha^2 G_F^2}{4 \pi\,m^3_\Sigma}
\nonumber\\
&\times& \left\{ \bigl[(2m_l^2 +
q^2)((m_p^{}-m_\Sigma^{})^2-q^2)(m_\Sigma^{} + m_p^{})^2 + 2 q^2\,
f(m_p^{}, m_\Sigma^{}, m_l, q^2,t)\bigr] \frac{|a|^2}{q^4} \right.
\nonumber\\
&&+\,\, \bigl[(2m_l^2 +
q^2)((m_p^{}+m_\Sigma^{})^2-q^2)(m_\Sigma^{} - m_p^{})^2 + 2 q^2\,
f(m_p^{}, m_\Sigma^{}, m_l, q^2,t) \bigr] \frac{|b|^2}{q^4}
\nonumber\\
&&+\,\,
\bigl[(2m_l^2+q^2)((m_p^{}-m_\Sigma^{})^2-q^2)-2f(m_p^{},m_\Sigma^{},m_l,q^2,t)\bigr]\,|c|^2
\nonumber\\
&&+\,\,
\bigl[(2m_l^2+q^2)((m_p^{}+m_\Sigma^{})^2-q^2)-2f(m_p^{},m_\Sigma^{},m_l,q^2,t)\bigr]\,|d|^2
\nonumber\\
&&+\,\,
2(m_\Sigma^{}+m_p^{})(2m_l^2+q^2)\bigl[(m_p^{}-m_\Sigma^{})^2-q^2)\bigr]\,
\frac{{\rm Re}\,({a c^*})}{q^2}
\nonumber\\
&&- \left. 2(m_\Sigma^{}-m_p^{}) (2m_l^2 + q^2) \bigl[(m_p^{}+
m_\Sigma^{})^2-q^2\bigr]\, \frac{{\rm Re}\,(b d^*)}{q^2}\right \}
\,\,,
\end{eqnarray}
where  \,$t=(p_\Sigma^{}-p_{\ell^-}^{})^2$\,  and
\begin{eqnarray}
f(m_p^{},m_\Sigma^{}, m_l,q^2,t) = m_l^4+ (m^2_p+m^2_\Sigma -q^2-
2t) m^2_l + m^2_p m^2_\Sigma - (m^2_p + m^2_\Sigma) t + (q^2+t)
t\;,\nonumber
\end{eqnarray}
with the integration intervals given by
\begin{eqnarray}
\begin{array}{c}   \displaystyle
t_{\rm max, min}^{}  \,\,=\,\, \mbox{$\frac{1}{2}$} \left[
m^2_\Sigma +m^2_p + 2 m^2_l -q^2 \pm \sqrt{1-\frac{4m^2_l}{q^2}}
\sqrt{(m^2_\Sigma-m^2_p-q^2)^2 -4m^2_p q^2} \right]   \,\,,
\vspace{2ex} \\   \displaystyle q^2_{\rm min}  \,\,=\,\,  4 m^2_l
\,\,,  \hspace{2em} q^2_{\rm max}  \,\,=\,\, (m_\Sigma^{} -
m_p^{})^2   \,\,.
\end{array}
\end{eqnarray}
It is worth mentioning that, since the form factors belong to the
\,$\Sigma^+\to p\gamma^*$\, amplitude, they do not depend on $t$.

\section{Imaginary parts of form factors in $\bm{\chi}$PT\label{imabcd}}

The chiral Lagrangian for the interactions of the lowest-lying
mesons and baryons is written down in terms of the lightest meson-octet and
baryon-octet fields, which are collected into  $3\times3$  matrices
$\varphi$  and  $B$,  respectively~\cite{Bijnens:1985kj}.
The mesons enter through the exponential  $\,\Sigma=\xi^2=\exp({\rm i}\varphi/f),\,$
where  $\,f=f_\pi^{}=92.4\rm\,MeV\,$  is the pion decay constant.
In the relativistic baryon~$\chi$PT, the lowest-order strong Lagrangian is given
by~\cite{Bijnens:1985kj}
\begin{eqnarray}   \label{Ls1}
{\cal L}_{\rm s}^{} &=&
\bigl\langle \bar{B}\, i\gamma^\mu
\bigl(\partial_\mu^{}B+\bigl[{\cal V}_\mu^{},B\bigr]\bigr) \bigr\rangle
+ m_0^{}\, \langle \bar{B}B \rangle
+ D\, \bigl\langle \bar{B}\gamma^\mu\gamma_5^{}\,
\bigl\{ {\cal A}_\mu, B \bigr\} \bigr\rangle
+ F\, \bigl\langle \bar{B}\gamma^\mu\gamma_5^{}\,
\bigl[ {\cal A}_\mu, B \bigr] \bigr\rangle   \,\,,
\hspace*{2em}
\end{eqnarray}
where  $\,\langle\cdots\rangle\equiv{\rm Tr}(\cdots)\,$  in flavor space,
$m_0^{}$  is the baryon mass in the chiral limit,
$\,{\cal V}^\mu=\frac{1}{2}\bigl(\xi\,\partial^\mu\xi^\dagger+\xi^\dagger\,\partial^\mu\xi\bigr)
+ \frac{i}{2}\,e A^\mu\bigl(\xi^\dagger Q\xi+\xi Q\xi^\dagger\bigr),\,$
and
$\,{\cal A}^\mu=\frac{i}{2}\bigl(\xi\,\partial^\mu\xi^\dagger-\xi^\dagger\,\partial^\mu\xi\bigr)
+ \frac{1}{2}\,e A^\mu\bigl(\xi^\dagger Q\xi-\xi Q\xi^\dagger\bigr),\,$
with  $A^\mu$  being the photon field and   $\,Q={\rm diag}(2,-1,-1)/3\,$  the quark-charge
matrix.\footnote{
Under a chiral transformation,  \,$\bar B\to U\bar B U^\dagger$,\, \,$B\to U B U^\dagger$,\,
\,${\cal V}^\mu\to U{\cal V}^\mu U^\dagger+i\partial^\mu U\, U^\dagger$,\,  and
\,${\cal A}^\mu\to U{\cal A}^\mu U^\dagger$,\,  where  $U$
is defined by  \,$\xi\to L\xi U^\dagger=U\xi R^\dagger$.}
The parameters $D$ and $F$  will enter our results below only through the combination
$\,D+F=1.26.\,$

\begin{figure}[tb]
\includegraphics{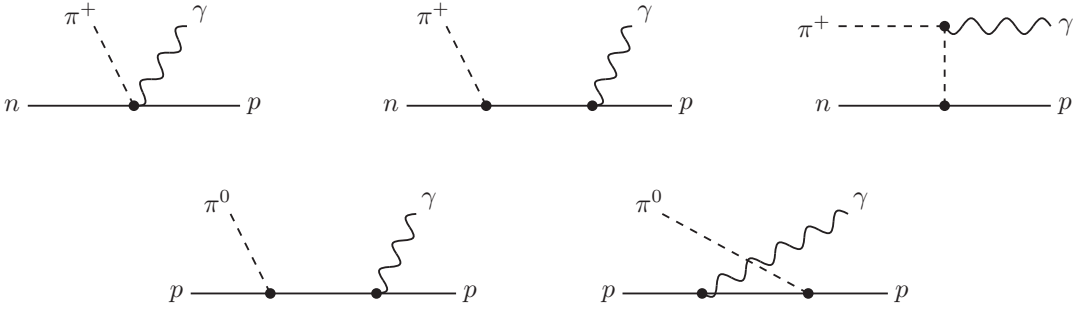}
\caption{Leading-order diagrams for  $\,N\pi\to p\gamma^*$ reactions. \label{NpiNg}}
\end{figure}

From  ${\cal L}_{\rm s}^{}$  we derive two sets of diagrams, shown in Fig.~\ref{NpiNg},
which represent the $\,N\pi\to p\gamma^*\,$  reactions involved in the unitarity calculation
of the imaginary parts of the form factors  $a$, $b$, $c$, and $d$.
It then follows from Fig.~\ref{fig_cut} that the first set of diagrams is associated with
the weak transition $\,\Sigma^{+}\to n\pi^{+}$,\,  and the second with  $\,\Sigma^{+}\to p\pi^{0}$.\,
Consequently,  we express our results as
\begin{eqnarray}   \label{imFF_r}
{\rm Im}\,{\cal F} &=& \frac{(D+F)m_{\pi^{+}}^{2}}{8\sqrt{2}\,\pi f_{\pi}}
\left(\tilde{\cal F}_+^{} + \frac{\tilde{\cal F}_0^{}}{\sqrt{2}}\right)
\hspace{2em}  \mbox{for  $\,\,{\cal F}=a,b,c,d$}  \,\,,
\end{eqnarray}
where  $\tilde{\cal F}_+^{}$  $\bigl(\tilde{\cal F}_0^{}\bigr)$
comes from the  $n\pi^+$  $(p\pi^0)$  contribution, and write them in terms of
the weak amplitudes  $\,A_+^{}=A_{n\pi^+}$,\,  $\,A_0^{}=A_{p\pi^0}$,\,
\,$B_+^{}=B_{n\pi^+}$,\, and  $\,B_0^{}=B_{p\pi^0}$\,  given in  Eq.~(\ref{ABnlhd}).
Working in the $\Sigma^+$  rest-frame, which implies that the
energies and momenta of the photon and proton in the final state
and of the pion in the intermediate are fixed by kinematics, we
define
\begin{eqnarray}
z_{+}^{}  \,\,=\,\,
\left(\frac{2 E_{\pi}^{} E_{\gamma}^{}-2 |\bm{p}_{\pi}^{}|\,|\bm{p}_{\gamma}^{}|-q^{2}}
           {2 E_{\pi}^{} E_{\gamma}^{}+2 |\bm{p}_{\pi}^{}|\,|\bm{p}_{\gamma}^{}|-q^{2}}\right)   \,\,,
\,\,&&\,\,
z_{0}^{}  \,\,=\,\,
\left(\frac{2 E_{\pi}^{} E_{p}^{}-2 |\bm{p}_{\pi}^{}|\,|\bm{p}_{p}^{}|-m_{\pi}^{2}}
           {2 E_{\pi}^{} E_{p}^{}+2 |\bm{p}_{\pi}^{}|\,|\bm{p}_{p}^{}|-m_{\pi}^{2}}\right)   \,\,.
\end{eqnarray}
The expression for $\tilde{\cal F}$  from each set of diagrams can then be written as
\begin{eqnarray}
\tilde{a}_{+,0}^{} &=&
\frac{B_{+,0}^{}\, m_N^{}}{2m_\Sigma^{2}\, |\bm{p}_{\gamma}^{}|}\,
\frac{\left[2 |\bm{p}_{\pi}^{}|\, |\bm{p}_{\gamma}^{}|\, f_{+,0}^{(a)}+\ln(z_{+,0}^{})\, g_{+,0}^{(a)}\right]}
     {\left[(m_\Sigma^{}-m_N^{})^{2}-q^{2}\right] \left[(m_\Sigma^{}+m_N^{})^{2}-q^{2}\right]^{2} }
\,\,, \nonumber \\
\tilde{b}_{+,0}^{} &=&
\frac{-A_{+,0}^{}\, m_N^{}}{2m_\Sigma^{2}\, |\bm{p}_{\gamma}^{}|}\,
\frac{\left[2 |\bm{p}_{\pi}^{}|\, |\bm{p}_{\gamma}^{}|\, f_{+,0}^{(b)}+\ln(z_{+,0}^{})\, g_{+,0}^{(b)}\right]}
     {\left[(m_\Sigma^{}-m_N^{})^{2}-q^{2}\right]^{2} \left[(m_\Sigma^{}+m_N^{})^{2}-q^{2}\right] }
\,\,, \nonumber \\
\tilde{c}_{+,0}^{} &=&
\frac{B_{+,0}^{}\, m_N^{}}{2m_\Sigma^{2}\, |\bm{p}_{\gamma}^{}|\,(m_N^{}-m_\Sigma^{})}\,
\frac{\left[2 |\bm{p}_{\pi}^{}|\, |\bm{p}_{\gamma}^{}|\, f_{+,0}^{(c)}+\ln(z_{+,0}^{})\, g_{+,0}^{(c)}\right]}
     {\left[(m_\Sigma^{}-m_N^{})^{2}-q^{2}\right] \left[(m_\Sigma^{}+m_N^{})^{2}-q^{2}\right]^{2} }
\,\,, \nonumber \\
\tilde{d}_{+,0}^{} &=&
\frac{-A_{+,0}^{}\, m_N^{}}{2m_\Sigma^{2}\, |\bm{p}_{\gamma}^{}|\,(m_\Sigma^{}+m_N^{})}\,
\frac{\left[2 |\bm{p}_{\pi}^{}|\, |\bm{p}_{\gamma}^{}|\, f_{+,0}^{(d)}+\ln(z_{+,0}^{})\, g_{+,0}^{(d)}\right]}
     {\left[(m_\Sigma^{}-m_N^{})^{2}-q^{2}\right]^{2} \left[(m_\Sigma^{}+m_N^{})^{2}-q^{2}\right]}  \,\,,
\end{eqnarray}
where
\begin{subequations}
\begin{eqnarray}
f_{+}^{(a)} &=&   m_N^{} m_\Sigma^5+\left(q^2+2 m_{\pi }^2+m_N^2\right) m_\Sigma^4
-m_N^{} \left(3 q^2-3 m_{\pi }^2+2 m_N^2\right) m_\Sigma^3
\nonumber \\
&&-\,  \left(q^4-5m_{\pi }^2 q^2+2 m_N^4+\left(q^2+m_{\pi }^2\right) m_N^2\right) m_\Sigma^2
\nonumber \\
&&+\,  m_N^{} \left(m_N^2-q^2\right) \left(2 q^2-3 m_{\pi }^2+m_N^2\right)m_\Sigma^{}
+ \left(q^2-m_N^2\right)^2 \left(m_N^2-m_{\pi }^2\right)
\,\,, \nonumber \\
g_{+}^{(a)} &=&   m_\Sigma^{} \left(m_N^{}
   q^6+\left(m_N^{} \left(2 m_N^{}-m_\Sigma^{}\right)
   \left(m_N^{}+m_\Sigma^{}\right)-m_{\pi }^2 \left(3
   m_N^{}+m_\Sigma^{}\right)\right) q^4\right.
\nonumber \\
&&+\, \left. m_{\pi }^2 \left(3
   m_{\pi }^2-4 m_N^2\right) \left(m_N^{}+m_\Sigma^{}\right)
   q^2+m_{\pi }^2 \left(m_N^{}-m_\Sigma^{}\right)^2
   \left(m_N^{}+m_\Sigma^{}\right)^3\right)
\,\,, \nonumber \\
f_{0}^{(a)} &=&   3 m_N^{} m_\Sigma^5-\left(q^2-2 m_{\pi }^2-3 m_N^2\right)m_\Sigma^4
-m_N^{} \left(4 m_N^2-3 \left(q^2+m_{\pi }^2\right)\right)m_\Sigma^3
\nonumber \\
&&+\, \left(q^4+5 m_{\pi }^2 q^2-4 m_N^4-\left(q^2+m_{\pi }^2\right) m_N^2\right) m_\Sigma^2
\nonumber \\
&&+\,
m_N^{} \left(m_N^2-q^2\right) \left(2 q^2-3 m_{\pi }^2+m_N^2\right) m_\Sigma^{}
+  \left(q^2-m_N^2\right)^2 \left(m_N^2-m_{\pi }^2\right)
\,\,, \nonumber \\
g_{0}^{(a)} &=& m_\Sigma^{} \left(-2 m_{\pi }^2
   m_\Sigma^{} q^4-\left(m_N^{}+m_\Sigma^{}\right) \left(3
   m_{\pi }^4-2 \left(3 m_N^2-2 m_\Sigma^{} m_N^{}+m_\Sigma^2\right) m_{\pi }^2 \right.\right.
\nonumber \\
&&+\,  \left.\left. m_N^{} \left(m_N^{}-m_{\Sigma }\right)^2 \left(3 m_N^{}+m_\Sigma^{}\right)\right)
   q^2+m_N^{} \left(m_N^{}-m_\Sigma^{}\right)^2 m_\Sigma^{}
   \left(m_N^{}+m_\Sigma^{}\right)^3\right)
\,\,,  \hspace*{2em}
\end{eqnarray}
\begin{eqnarray}
f_{+}^{(b)} &=&   m_N^{} m_\Sigma^5-\left(q^2+2 m_{\pi }^2+m_N^2\right) m_\Sigma^4
-m_N^{} \left(3 q^2-3 m_{\pi }^2+2 m_N^2\right) m_\Sigma^3
\nonumber \\
&&+\,  \left(q^4-5 m_{\pi }^2 q^2+2 m_N^4+\left(q^2+m_{\pi }^2\right) m_N^2\right) m_\Sigma^2
\nonumber \\
&&+\,
m_N^{} \left(m_N^2-q^2\right) \left(2 q^2-3 m_{\pi }^2+m_N^2\right) m_\Sigma^{}
+  \left(q^2-m_N^2\right)^2 \left(m_{\pi }^2-m_N^2\right)
\,\,, \nonumber \\
g_{+}^{(b)} &=&   - m_\Sigma^{} \left(-m_N^{}
   q^6+\left(\left(3 m_N^{}-m_\Sigma^{}\right) m_{\pi }^2+m_N^{}
   \left(-2 m_N^2+m_\Sigma^{} m_N^{}+m_{\Sigma }^2\right)\right) q^4\right.\nonumber\\
&&-\,  \left. m_{\pi }^2 \left(3 m_{\pi }^2-4
   m_N^2\right) \left(m_N^{}-m_\Sigma^{}\right) q^2-m_{\pi
   }^2 \left(m_N^{}-m_\Sigma^{}\right)^3 \left(m_N^{}+m_{\Sigma }\right)^2\right)
\,\,, \nonumber \\
f_{0}^{(b)} &=&   3 m_N^{} m_\Sigma^5+\left(q^2-2 m_{\pi }^2-3 m_N^2\right) m_\Sigma^4
+m_N^{} \left(3 \left(q^2+m_{\pi }^2\right)-4 m_N^2\right) m_\Sigma^3
\nonumber \\
&&-\, \left(q^4+5 m_{\pi }^2 q^2-4 m_N^4-\left(q^2+m_{\pi }^2\right) m_N^2\right) m_\Sigma^2
\nonumber \\
&&+\, m_N^{} \left(m_N^2-q^2\right) \left(2 q^2-3 m_{\pi }^2+m_N^2\right) m_\Sigma^{}
+ \left(q^2-m_N^2\right)^2 \left(m_{\pi }^2-m_N^2\right)
\,\,, \nonumber \\
g_{0}^{(b)} &=&   -  m_\Sigma^{} \left(-2 m_{\pi }^2
   m_\Sigma^{} q^4+\left(m_N^{}-m_\Sigma^{}\right) \left(3
   m_{\pi }^4-2 \left(3 m_N^2+2 m_\Sigma^{} m_N^{}+m_{\Sigma }^2\right) m_{\pi }^2\right.\right.\nonumber\\
&&+\,  \left.\left. m_N^{} \left(3 m_N^{}-m_{\Sigma }\right) \left(m_N^{}+m_\Sigma^{}\right)^2\right) q^2+m_N^{}
   \left(m_N^{}-m_\Sigma^{}\right)^3 m_\Sigma^{}
   \left(m_N^{}+m_\Sigma^{}\right)^2\right)
\,\,, \hspace*{2em}
\end{eqnarray}
\begin{eqnarray}
f_{+}^{(c)} &=&   m_{\pi }^2 \left(8 m_\Sigma^4+5 m_N^{} m_\Sigma^3
-\left(3 q^2+m_N^2\right) m_\Sigma^2+3 m_N^{} \left(m_N^2-q^2\right) m_\Sigma^{}
+\left(q^2-m_N^2\right)^2\right)
\nonumber \\
&&-\, \left(m_N^{}-m_\Sigma^{}\right) \left(-m_N^{} m_\Sigma^4
+\left(q^2-2 m_N^2\right) m_\Sigma^3-4 q^2 m_N^{} m_\Sigma^2
\vphantom{|_|^|}
\right. \nonumber \\
&&-\, \left.
\left(q^4+m_N^2 q^2-2 m_N^4\right) m_\Sigma^{} \right.
+  \left. m_N^{} \left(q^2-m_N^2\right)^2\right)
\,\,, \nonumber \\
g_{+}^{(c)} &=&   -  \left(m_N^{}-m_\Sigma^{}\right)
   m_\Sigma^{} \left(m_N^{} \left(2 m_N^{}+m_\Sigma^{}\right)
   q^4+\left(m_{\pi }^4-2 \left(3 m_N^2+2 m_\Sigma^{}
   m_N^{}+m_\Sigma^2\right) m_{\pi }^2\right.\right.\nonumber\\
&&+\,  \left.\left.m_N^{}
   \left(m_N^{}-m_\Sigma^{}\right) \left(m_N^{}+m_{\Sigma }\right)^2\right) q^2+2 m_{\pi }^2
   \left(m_N^{}+m_{\Sigma }\right)^2 \left(m_{\pi }^2+m_\Sigma^{} \left(m_{\Sigma }-m_N^{}\right)\right)\right)
\,\,, \nonumber \\
f_{0}^{(c)} &=&   m_{\pi }^2 \left(8 m_\Sigma^4+5 m_N^{} m_\Sigma^3
-\left(3 q^2+m_N^2\right) m_\Sigma^2+3 m_N^{} \left(m_N^2-q^2\right) m_\Sigma^{}
+\left(q^2-m_N^2\right)^2\right)
\nonumber \\
&&-\, \left(m_N^{}-m_\Sigma^{}\right)^2 \left(2 m_\Sigma^4-m_N^{} m_\Sigma^3
-\left(3 q^2+m_N^2\right) m_\Sigma^2+3 m_N^{} \left(m_N^2-q^2\right) m_\Sigma^{}
\vphantom{|_|^|}
\right. \nonumber \\
&&+\, \left.
\left(q^2-m_N^2\right)^2\right)
\,\,, \nonumber \\
g_{0}^{(c)} &=&
\left(m_N^{}-m_\Sigma^{}\right) m_{\Sigma } \left(\left(m_{\pi }^4-\left(2 m_N^2-4 m_\Sigma^{}
   m_N^{}-2 m_\Sigma^2\right) m_{\pi }^2
+ m_N^{}
   \left(m_N^{}-m_\Sigma^{}\right)^2 \left(m_N^{}+m_{\Sigma }\right)\right) q^2\right.
\nonumber\\
&&+\,  \left. \left(m_N^{}+m_\Sigma^{}\right)^2
   \left(2 m_{\pi }^4-2 \left(2 m_N^2-m_\Sigma^{}
   m_N^{}+m_\Sigma^2\right) m_{\pi }^2
+ m_N^{}
   \left(m_N^{}-m_\Sigma^{}\right)^2 \left(2 m_N^{}-m_{\Sigma }\right)\right)\right)   \,\,,
\nonumber \\
\end{eqnarray}
\begin{eqnarray}
f_{+}^{(d)} &=&   \left(m_N^{}+m_\Sigma^{}\right) \left(-m_N^{} m_\Sigma^4
-\left(q^2-2 m_N^2\right) m_\Sigma^3-4 q^2 m_N^{} m_\Sigma^2
\vphantom{|_|^|}
\right. \nonumber \\
&&+\,  \left.
\left(q^4+m_N^2 q^2-2 m_N^4\right) m_\Sigma^{} \right.
+ \left. m_N^{} \left(q^2-m_N^2\right)^2\right)
\nonumber \\
&&-\,\,
m_{\pi }^2 \left(8 m_\Sigma^4-5 m_N^{} m_\Sigma^3-\left(3 q^2+m_N^2\right) m_\Sigma^2
+ 3 m_N^{} \left(q^2-m_N^2\right) m_\Sigma^{}
+ \left(q^2-m_N^2\right)^2\right)
\,\,, \nonumber \\
g_{+}^{(d)} &=&   m_\Sigma^{} \left(m_N^{}+m_{\Sigma }\right) \left(-m_N^{} \left(2 m_N^{}-m_\Sigma^{}\right)
   q^4-\left(m_{\pi }^4-2 \left(3 m_N^2-2 m_\Sigma^{}
   m_N^{}+m_\Sigma^2\right) m_{\pi }^2\right.\right.
\nonumber\\
&&+\,  \left.\left. m_N^{}
   \left(m_N^{}-m_\Sigma^{}\right)^2 \left(m_N^{}+m_{\Sigma }\right)\right) q^2
- 2 m_{\pi }^2 \left(m_N^{}-m_{\Sigma }\right)^2 \left(m_{\pi }^2+m_\Sigma^{}
   \left(m_N^{}+m_\Sigma^{}\right)\right)\right)
\,\,, \nonumber \\
f_{0}^{(d)} &=&   \left(m_N^{}+m_\Sigma^{}\right)^2 \left(2 m_\Sigma^4
+m_N^{} m_\Sigma^3-\left(3 q^2+m_N^2\right) m_\Sigma^2
\vphantom{|_|^|}
+ 3 m_N^{} \left(q^2-m_N^2\right) m_\Sigma^{}
+\left(q^2-m_N^2\right)^2\right)
\nonumber \\
&&-\,\,  m_{\pi }^2 \left(8 m_\Sigma^4-5 m_N^{} m_\Sigma^3-\left(3 q^2+m_N^2\right) m_\Sigma^2
+3 m_N^{} \left(q^2-m_N^2\right) m_\Sigma^{}+\left(q^2-m_N^2\right)^2\right)
\,\,, \nonumber \\
g_{0}^{(d)} &=&  m_\Sigma^{} \left(m_N^{}+m_{\Sigma }\right) \left(\left(m_{\pi }^4-2 \left(m_N^2+2
   m_\Sigma^{} m_N^{}-m_\Sigma^2\right) m_{\pi }^2
+ m_N^{}
   \left(m_N^{}-m_\Sigma^{}\right) \left(m_N^{}+m_{\Sigma }\right)^2\right) q^2\right.\nonumber \\
&&+\,  \left. \left(m_N^{}-m_\Sigma^{}\right)^2
   \left(2 m_{\pi }^4-2 \left(2 m_N^2+m_\Sigma^{}
   m_N^{}+m_\Sigma^2\right) m_{\pi }^2
\right.\right. \nonumber \\
&&+\,  \left.\left.
m_N^{}
   \left(m_N^{}+m_\Sigma^{}\right)^2 \left(2 m_N^{}+m_{\Sigma }\right)\right)\right)   \,\,.
\end{eqnarray}
\end{subequations}
In our numerical computations,  $\,m_\Sigma^{}=m_{\Sigma^+}^{},\,$
$\,m_N^{}=\frac{1}{2}\bigl(m_p^{}+m_n^{}\bigr),\,$
$\,m_\pi^{}=\frac{1}{3}\bigl(2m_{\pi^+}^{}+m_{\pi^0}^{}\bigr),\,$
the numbers being from Ref.~\cite{pdg}.

\begin{figure}[b]
\includegraphics{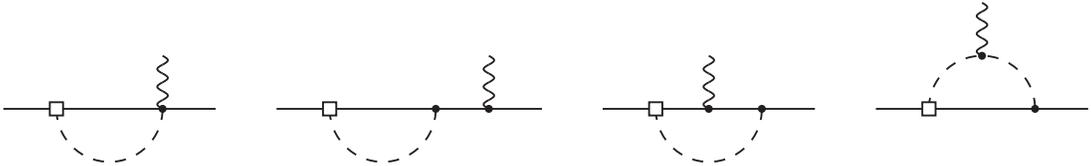}
\caption{Diagrams for imaginary part of  \,$\Sigma^+\to p\gamma^*$\, amplitude.\label{loops}}
\end{figure}

In heavy baryon $\chi$PT~\cite{Jenkins:1991ne}, the relevant Lagrangian can be found in
Ref.~\cite{Jenkins:1992ab},  and the weak radiative and nonleptonic amplitudes in
Eqs.~(\ref{M_BBg})  and~(\ref{M_SNpi}) become, respectively,
\begin{eqnarray}
{\cal M}(B_i\to B_f\gamma^*)  &=&
-e G_{\rm F}^{}\, \bar{B}_f\, \Bigl[
2 \bigl(S\cdot q\,S^\mu-S^\mu\,S\cdot q\bigr)a+2\bigl(S\cdot q\,v^\mu-S^\mu\,v\cdot q\bigr) b
\Bigr] \, B_i\, \varepsilon_\mu^*
\nonumber \\ &&
-\,\,
e G_{\rm F}^{}\, \bar{B}_f\, \Bigl[
\bigl(q^2\,v^\mu-q^\mu\, v\cdot q\bigr) c+2 \bigl(q^2\,S^\mu-q^\mu\,S\cdot q\bigr) d
\Bigr] \, B_i\, \varepsilon_\mu^*   \,\,,
\end{eqnarray}
\begin{eqnarray}
{\cal M}(\Sigma^+\to N\pi)  &=&
i G_{F}^{} m_{\pi^+}^2\, \bar{N} \left( A_{N\pi}^{}
+ 2S\cdot p_\pi^{}\, \frac{B_{N\pi}^{}}{2m_\Sigma^{}} \right) \Sigma   \,\,,
\end{eqnarray}
where  $v$  is the baryon four-velocity and  $S$ is the baryon spin operator.
Following Ref.~\cite{Jenkins:1992ab},  to obtain the imaginary parts of the form factors
we evaluate the loop diagrams displayed in Fig.~\ref{loops}.
In the heavy-baryon approach, only the diagrams with the  $\,\Sigma^+\to n\pi^+\,$
transition yield nonzero contributions to the leading-order imaginary parts.
The results are
\begin{subequations}   \label{imFF_hb}
\begin{eqnarray}
{\rm Im}\,a  &=&
\frac{(D+F)\,m_{\pi^+}^2}{8\sqrt2\,\pi f_\pi^{}}\, \frac{B_{n\pi^+}^{}}{2m_\Sigma^{}}
\left\{ \sqrt{\Delta^2-m_\pi^2}\Biggl( 1+\frac{\frac{1}{2}\, q^2}{\Delta^2-q^2}\Biggr)
\right.
\nonumber \\ &&
\left.
+\,\,
\frac{q^4+4 m_\pi^2\bigl(\Delta^2-q^2\bigr)}{4\bigl(\Delta^2-q^2\bigr)^{3/2}}
 \ln \left[ \frac{2\Delta^2-q^2-2\sqrt{\Delta^2-m_\pi^2}\sqrt{\Delta^2-q^2}}
           {\sqrt{q^4+4m_\pi^2\bigl(\Delta^2-q^2\bigr)}} \right]
\right\}   \,\,,
\end{eqnarray}
\begin{eqnarray}
{\rm Im}\,b   &=&
\frac{(D+F)\,m_{\pi^+}^2}{8\sqrt2\,\pi f_\pi^{}}\, A_{n\pi^+}^{}
\left\{ \frac{-\Delta\, \sqrt{\Delta^2-m_\pi^2}}{\Delta^2-q^2}
       \Biggl( 1-\frac{\frac{3}{2}\, q^2}{\Delta^2-q^2}\Biggr)
\right.
\nonumber \\ &&
\left.
+\,\,
\Delta\, \frac{3q^4+4 m_\pi^2\bigl(\Delta^2-q^2\bigr)}
                {4\bigl(\Delta^2-q^2\bigr)^{5/2}}
 \ln \left[ \frac{2\Delta^2-q^2-2\sqrt{\Delta^2-m_\pi^2}\sqrt{\Delta^2-q^2}}
           {\sqrt{q^4+4m_\pi^2\bigl(\Delta^2-q^2\bigr)}} \right]
\right\}   \,\,,
\hspace*{2em}
\end{eqnarray}
\begin{eqnarray}
{\rm Im}\,c  &=&
\frac{(D+F)\,m_{\pi^+}^2}{8\sqrt2\,\pi f_\pi^{}}\, \frac{B_{n\pi^+}^{}}{2m_\Sigma^{}}
\left\{ \sqrt{\Delta^2-m_\pi^2}\, \frac{\Delta^2-2 m_\pi^2}{\Delta\,\bigl(\Delta^2-q^2\bigr)}
\right.
\nonumber \\ &&
\left.
+\,\,
\frac{\Delta\, \bigl(q^2-2 m_\pi^2\bigr)}{2\bigl(\Delta^2-q^2\bigr)^{3/2}}\,
\ln \left[ \frac{2\Delta^2-q^2-2\sqrt{\Delta^2-m_\pi^2}\sqrt{\Delta^2-q^2}}
        {\sqrt{q^4+4m_\pi^2\bigl(\Delta^2-q^2\bigr)}} \right]
\right\}   \,\,,
\end{eqnarray}
\begin{eqnarray}
{\rm Im}\,d  &=&
\frac{(D+F)\,m_{\pi^+}^2}{8\sqrt2\,\pi f_\pi^{}}\, A_{n\pi^+}^{} \left\{
\sqrt{\Delta^2-m_\pi^2}\, \frac{\frac{3}{2}\, q^2}{\bigl(\Delta^2-q^2\bigr)^2}
\right.
\nonumber \\ &&
\left.
+\,\,
\frac{q^4+2q^2\Delta^2+4 m_\pi^2\bigl(\Delta^2-q^2\bigr)}
     {4\bigl(\Delta^2-q^2\bigr)^{5/2}}\,
\ln \left[ \frac{2\Delta^2-q^2-2\sqrt{\Delta^2-m_\pi^2}\sqrt{\Delta^2-q^2}}
        {\sqrt{q^4+4m_\pi^2\bigl(\Delta^2-q^2\bigr)}} \right]
\right\}   \,\,,
\hspace*{3em}
\end{eqnarray}
\end{subequations}
where  $\,\Delta=m_\Sigma^{}-m_N^{}.\,$
We have checked that these formulas can be reproduced from the relativistic results
in Eq.~(\ref{imFF_r})  by expanding the latter in terms of  $\Delta/m_\Sigma^{}$,
$\sqrt{q^2}/m_\Sigma^{}$,  and  $m_\pi^{}/m_\Sigma^{}$  and  keeping the leading nonzero
terms.

\section{Real parts of $\bm{c(q^2)}$ and $\bm{d(q^2)}$\label{recd}}

\begin{figure}[b]
\includegraphics{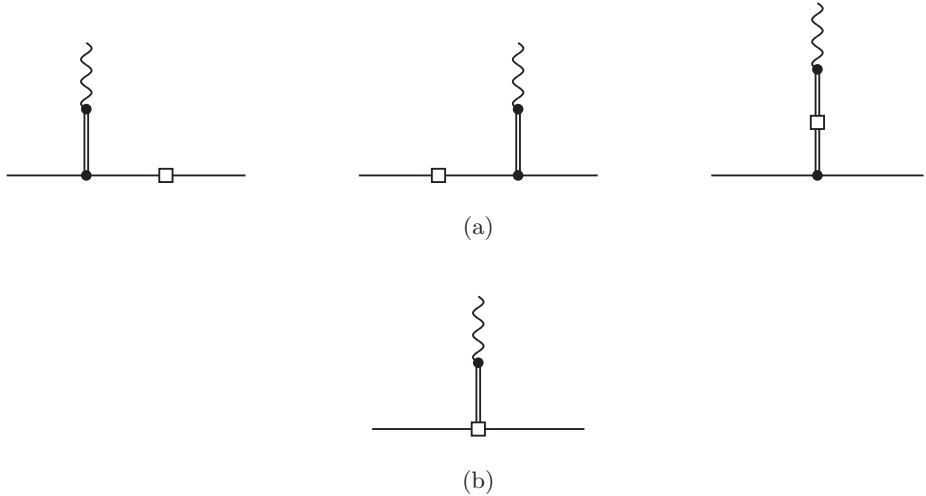}
\caption{\label{cd-poles}%
Pole diagrams contributing to the $c$ and $d$ amplitudes.
A single line (double line) denotes a baryon (vector meson) field,
and a solid dot (hollow square) represents a strong (weak) vertex.
}\end{figure}

Vector mesons can contribute to $c$ via the pole diagrams shown in Fig.~\ref{cd-poles}(a).
The strong vertices in the diagrams come from the
Lagrangian~\cite{Ecker:1989yg,Kubis:2000zd}
\begin{eqnarray}   \label{Ls'}
{\cal L}_{\rm s}'  &=&
{\cal G}_D^{}\,\bigl\langle\bar{B}\,\gamma^\mu\,\bigl\{{\sf V}_\mu^{},B\bigr\}\bigr\rangle
+ {\cal G}_F^{}\,\bigl\langle\bar{B}\,\gamma^\mu\,\bigl[{\sf V}_\mu^{},B\bigr]\bigr\rangle
+ {\cal G}_0^{}\,\bigl\langle\bar{B}\gamma^\mu B\bigr\rangle\,
\bigl\langle{\sf V}_\mu^{}\bigr\rangle
\nonumber \\ &&
-\,\,
\mbox{$\frac{1}{2}$}\, e f_{\sf V}^{}\,
\bigl\langle \bigl(D^\mu{\sf V}^\nu-D^\nu{\sf V}^\mu\bigr)
\bigl( \xi^\dagger Q\xi+\xi Q\xi^\dagger\bigr) \bigr\rangle\,
\bigl( \partial_\mu^{}A_\nu^{}-\partial_\nu^{}A_\mu^{}\bigr)   \,\,,
\end{eqnarray}
with  $\,{\sf V}=\frac{1}{2}\lambda_3^{}\rho^0+\cdots$\,\,  containing the nonet of
vector-meson fields and
$\,D^\mu{\sf V}^\nu=\partial^\mu{\sf V}^\nu+\bigl[{\cal V}^\mu,{\sf V}^\nu\bigr]$,\footnote{
Under a chiral transformation,  \,${\sf V}\to U{\sf V}U^\dagger$\, and
\,$D^\mu{\sf V}^\nu\to U D^\mu{\sf V}^\nu U^\dagger$.}
whereas the weak vertices arise from
\begin{eqnarray}   \label{Lw}
{\cal L}_{\rm w}^{}  &=&  G_{\rm F}^{}m_{\pi^+}^2 \left(
h_D^{}\, \bigl\langle \bar{B}\, \bigl\{ \xi^\dagger h \xi, B \bigr\} \bigr\rangle
+ h_F^{}\, \bigl\langle \bar{B}\, \bigl[ \xi^\dagger h \xi, B \bigr] \bigr\rangle
\,+\,
h_{\sf V}^{}\, \bigl\langle h\, \xi{\sf V}^\mu{\sf V}_\mu^{}\xi^\dagger\bigr\rangle
\right)
\,\,+\,\,  {\rm H.c.}   \,\,,
\end{eqnarray}
with  $h$ being a 3$\times$3-matrix  having elements
$\,h_{kl}^{}=\delta_{k2}^{}\delta_{3l}^{}\,$  which selects out  $\,s\to d\,$
transitions.
The relevant parameters in ${\cal L}_{\rm s}'$ are \,${\cal G}_D^{}=-13.9$\, and
\,${\cal G}_F^{}=17.9$\, from a recent dispersive
analysis~\cite{Kubis:2000zd,Mergell:1995bf},\footnote{Although ${\cal G}_0^{}$ does
not appear in our results, it enters the extraction of ${\cal G}_{D,F}^{}$.
Writing the $pp\sf V$ part of  ${\cal L}_{\rm s}'$ as
\,$\frac{1}{2}\bar{p}\gamma^\mu p\, \bigl(g_{\rho NN}^{}\rho_\mu^0
+ g_{\omega NN}^{}\omega_\mu^{} + g_{\phi NN}^{}\phi_\mu^{}\bigr)$,\,
we have  \,$g_{\rho NN}^{}={\cal G}_D^{}+{\cal G}_F^{}=4.0$,\,
\,$g_{\omega NN}^{}={\cal G}_D^{}+{\cal G}_F^{}+2{\cal G}_0^{}=41.8$,\, and
\,$g_{\phi NN}^{}=\sqrt2\, \bigl({\cal G}_D^{}-{\cal G}_F^{}+{\cal G}_0^{}\bigr)=-18.3$,\,
where the numbers are from  Ref.~\cite{Kubis:2000zd,Mergell:1995bf}.
}
and  \,$f_{\sf V}^{}=0.201$\,  from  $\,\rho^0\to e^+ e^-\,$  rate~\cite{pdg},
while those in  ${\cal L}_{\rm w}^{}$ are
$\,h_D^{}=-72\,{\rm MeV}\,$  and  $\,h_F^{}=179\,{\rm MeV}$\, extracted at tree level from
S-wave hyperon nonleptonic decays~\cite{AbdEl-Hady:1999mj}, but  $h_{\sf V}^{}$
cannot be determined directly from data.
To estimate  $h_{\sf V}^{}$,  we use the SU(6$)_w^{}$ relation
$\,\bigl\langle\pi^0\bigr|{\cal H}_{\rm w}^{}\bigl|\bar{K}^0\bigr\rangle
= \bigl\langle\rho^0\bigr|{\cal H}_{\rm w}^{}\bigl|\bar{K}^{*0}\bigr\rangle\,$
derived in Ref.~\cite{Dubach:1996dg}.
Thus, employing the weak chiral Lagrangian
$\,{\cal L}_{\rm w}^\varphi=\gamma_8^{}\,f^2\, \bigl\langle h\,
\partial^\mu\Sigma\,\partial_\mu^{}\Sigma^\dagger \bigr\rangle  +{\rm H.c.},\,$
with  $\,\gamma_8^{}=7.8\times10^{-8}\,$  from  $\,K\to\pi\pi\,$  data,  we find
$\,h_{\rm V}^{}=-4\gamma_8^{}\, m_K^2/\bigl(G_{\rm F}^{}m_{\pi^+}^2\bigr)=-0.34{\rm\,GeV}^2$.\,
Putting things together and adopting ideal $\omega$-$\phi$ mixing, we then obtain
\begin{eqnarray}   \label{rec}
{\rm Re}\,c  &=&
\frac{f_{\sf V}^{}\, \bigl({\cal G}_D^{}-{\cal G}_F^{}\bigr)\, m_{\pi^+}^2\,
      \bigl(h_D^{}-h_F^{}\bigr)}{6 \bigl(m_\Sigma^{}-m_N^{}\bigr)}
\left(\frac{3}{q^2-m_\rho^2}-\frac{1}{q^2-m_\omega^2}
      - \frac{2}{q^2-m_\phi^2} \right)
\nonumber \\ && +\,\,
\frac{f_{\sf V}^{}\,\bigl({\cal G}_D^{}-{\cal G}_F^{}\bigr)\,m_{\pi^+}^2\,h_{\sf V}^{}}
     {12 \bigl(q^2-m_{K^*}^2\bigr)}
\left(\frac{3}{q^2-m_\rho^2}-\frac{1}{q^2-m_\omega^2}
      + \frac{2}{q^2-m_\phi^2} \right)   \,\,.
\end{eqnarray}

The form factor $d$ can receive vector-meson contributions from the parity-violating Lagrangian
\begin{eqnarray}   \label{Lw^PV}
{\cal L}_{\rm w}'  \,\,=\,\,
G_{\rm F}^{}m_{\pi^+}^2\, h_{\rm PV}^{}\, \bigl\langle h\,
\xi\, \bigl\{ \bigl[ \bar{B},\gamma^\mu\gamma_5^{}{ B}\bigr],{\sf V}_\mu^{} \bigr\}\,
\xi^\dagger \bigr\rangle
\,\,+\,\,  {\rm H.c.}   \,\,,
\end{eqnarray}
which are represented by the diagram in Fig.~\ref{cd-poles}(b).
The parameter  $h_{\rm PV}^{}$  also cannot be fixed directly from data,
and so we estimate it by adopting again the SU(6$)_w^{}$ results of
Ref.~\cite{Dubach:1996dg} to be  \,$h_{\rm PV}^{}=2.41.$\,
It follows that
\begin{eqnarray}   \label{red}
{\rm Re}\,d  &=&
\frac{f_{\sf V}^{}\, m_{\pi^+}^2\, h_{\rm PV}^{}}{6}
\left(\frac{3}{q^2-m_\rho^2}-\frac{1}{q^2-m_\omega^2}
      + \frac{2}{q^2-m_\phi^2} \right)   \,\,.
\end{eqnarray}


\begin{thebibliography}{99}

%\cite{Park:2005ek}
\bibitem{Park:2005ek}
  H.~Park {\it et al.}  [HyperCP Collaboration],
  %``Evidence for the decay Sigma+ $\to$ p mu+ mu-,''
  Phys.\ Rev.\ Lett.\  {\bf 94}, 021801 (2005)
  [arXiv:hep-ex/0501014].
  %%CITATION = HEP-EX 0501014;%%

%\cite{Bergstrom:1987wr}
\bibitem{Bergstrom:1987wr}
  L.~Bergstrom, R.~Safadi, and P.~Singer,
  %``Phenomenology Of Sigma+ $\to$ P Lepton+ Lepton- And The Structure Of The
  %Weak Nonleptonic Hamiltonian,''
  Z.\ Phys.\ C {\bf 37}, 281 (1988).
  %%CITATION = ZEPYA,C37,281;%%

%\cite{Buchalla:1995vs}
\bibitem{Buchalla:1995vs}
G.~Buchalla, A.J.~Buras, and M.E.~Lautenbacher,
%``Weak Decays Beyond Leading Logarithms,''
Rev.\ Mod.\ Phys.\  {\bf 68}, 1125 (1996) [arXiv:hep-ph/9512380].
%%CITATION = HEP-PH 9512380;%%

%\cite{He:1999ik}
\bibitem{He:1999ik}
  X.G.~He and G.~Valencia,
  %``Constraints on s $\to$ d gamma from radiative hyperon and kaon decays,''
  Phys.\ Rev.\ D {\bf 61}, 075003 (2000)
  [arXiv:hep-ph/9908298].
  %%CITATION = HEP-PH 9908298;%%

%\cite{Shifman:1976de}
\bibitem{Shifman:1976de}
  M.A.~Shifman, A.I.~Vainshtein, and V.I.~Zakharov,
  %``On The Weak Radiative Decays (Effects Of Strong Interactions At Short
  %Distances),''
  Phys.\ Rev.\ D {\bf 18}, 2583 (1978)
  [Erratum-ibid.\ D {\bf 19}, 2815 (1979)].
  %%CITATION = PHRVA,D18,2583;%%

\bibitem{ckm}
N.~Cabibbo, Phys.\ Rev.\ Lett.\  {\bf 10}, 531 (1963);
%%CITATION = PRLTA,10,531;%%
M.~Kobayashi and T.~Maskawa, Prog.\ Theor.\ Phys.\  {\bf 49}, 652 (1973).
%%CITATION = PTPKA,49,652;%%

%\cite{Donoghue:1992dd}
\bibitem{Donoghue:1992dd}
J.F.~Donoghue, E.~Golowich, and B.R.~Holstein, {\it Dynamics of the
Standard Model} (Cambridge University Press, Cambridge, 1992).
  %``Dynamics of the standard model,''
%  Camb.\ Monogr.\ Part.\ Phys.\ Nucl.\ Phys.\ Cosmol.\  {\bf 2}, 1 (1992).
  %%CITATION = CMPCE,2,1;%%


%\cite{Eidelman:2004wy}
\bibitem{pdg}
  S.~Eidelman {\it et al.}  [Particle Data Group],
  %``Review of particle physics,''
  Phys.\ Lett.\ B {\bf 592}, 1 (2004).
  %%CITATION = PHLTA,B592,1;%%


%\cite{Neufeld:1992hb}
\bibitem{Neufeld:1992hb}
  H.~Neufeld,
  %``Weak radiative baryon decays in chiral perturbation theory,''
  Nucl.\ Phys.\ B {\bf 402}, 166 (1993).
  %%CITATION = NUPHA,B402,166;%%

%\cite{Jenkins:1992ab}
\bibitem{Jenkins:1992ab}
  E.~Jenkins, M.E.~Luke, A.V.~Manohar, and M.J.~Savage,
  %``Weak radiative hyperon decays in chiral perturbation theory,''
  Nucl.\ Phys.\ B {\bf 397}, 84 (1993)
  [arXiv:hep-ph/9210265].
  %%CITATION = HEP-PH 9210265;%%

%\cite{Bos:1996ig}
\bibitem{Bos:1996ig}
  J.W.~Bos {\it et al.}, %D.Chang, S.C.~Lee, Y.~C.~Lin and H.~H.~Shih,
  %``Hyperon weak radiative decays in chiral perturbation theory,''
  Phys.\ Rev.\ D {\bf 54}, 3321 (1996)
  [arXiv:hep-ph/9601299];
  %%CITATION = HEP-PH 9601299;%%
  %``Heavy-baryon chiral perturbation theory and reparametrization
  %invariance,''
  {\it ibid.} {\bf 57}, 4101 (1998)
  [arXiv:hep-ph/9611260].
  %%CITATION = HEP-PH 9611260;%%


\bibitem{ang}
G.~Ang {\it et al.}, Z. Phys. {\bf 228}, 151 (1969).
%%CITATION = ZEPYA,228,151;%%


\bibitem{Bijnens:1985kj}
  J.~Bijnens, H.~Sonoda, and M.B.~Wise,
  %``On The Validity Of Chiral Perturbation Theory For Weak Hyperon Decays,''
  Nucl.\ Phys.\ B {\bf 261}, 185 (1985).
  %%CITATION = NUPHA,B261,185;%%


%\cite{Jenkins:1991ne}
\bibitem{Jenkins:1991ne}
E.~Jenkins and A.V.~Manohar, Phys.\ Lett.\ B {\bf 255}, 558 (1991);
%%CITATION = PHLTA,B255,558;%%
in {\it Effective Field Theories of the Standard Model},
edited by U.-G. Meissner (World Scientific, Singapore, 1992).
%``Baryon chiral perturbation theory,'' UCSD-PTH-91-30
%\href{http://www.slac.stanford.edu/spires/find/hep/www?r=ucsd-pth-91-30}{SPIRES entry}
%{\it Talk presented at the Workshop on Effective Field Theories of the Standard Model,
%Dobogoko, Hungary, Aug 1991}


%\cite{Ecker:1989yg}
\bibitem{Ecker:1989yg}
  G.~Ecker {\it et al.}, %J.~Gasser, H.~Leutwyler, A.~Pich and E.~de Rafael,
  %``Chiral Lagrangians For Massive Spin 1 Fields,''
  Phys.\ Lett.\ B {\bf 223}, 425 (1989);
  %%CITATION = PHLTA,B223,425;%%
%\cite{Borasoy:1995ds}
%\bibitem{Borasoy:1995ds}
  B.~Borasoy and U.G.~Meissner,
  %``Chiral Lagrangians for Baryons coupled to massive Spin-1 Fields,''
  Int.\ J.\ Mod.\ Phys.\ A {\bf 11}, 5183 (1996)
  [arXiv:hep-ph/9511320].
  %%CITATION = HEP-PH 9511320;%%

%\cite{Kubis:2000zd}
\bibitem{Kubis:2000zd}
  B.~Kubis and U.G.~Meissner,
  %``Low energy analysis of the nucleon electromagnetic form factors,''
  Nucl.\ Phys.\ A {\bf 679}, 698 (2001)
  [arXiv:hep-ph/0007056].
  %%CITATION = HEP-PH 0007056;%%

%\cite{Mergell:1995bf}
\bibitem{Mergell:1995bf}
  P.~Mergell, U.G.~Meissner, and D.~Drechsel,
  %``Dispersion theoretical analysis of the nucleon electromagnetic
  %form-factors,''
  Nucl.\ Phys.\ A {\bf 596}, 367 (1996)
  [arXiv:hep-ph/9506375];
  %%CITATION = HEP-PH 9506375;%%

\bibitem{AbdEl-Hady:1999mj}
A.~Abd El-Hady and J.~Tandean,
%``Hyperon nonleptonic decays in chiral perturbation theory reexamined,''
Phys.\ Rev.\ D {\bf 61}, 114014 (2000)
[arXiv:hep-ph/9908498].
%%CITATION = HEP-PH 9908498;%%


%\cite{Dubach:1996dg}
\bibitem{Dubach:1996dg}
  J.F.~Dubach, G.B.~Feldman, and B.R.~Holstein,
  %``Theory of weak hypernuclear decay,''
  Annals Phys.\  {\bf 249}, 146 (1996)
  [arXiv:nucl-th/9606003].
  %%CITATION = NUCL-TH 9606003;%%


\end{thebibliography}
\end{document}